# High temperature metamaterial TeraHertz quantum detector

Mathieu Jeannin,[1] Thomas Bonazzi,[1] Djamal Gacemi,[1] Angela Vasanelli,[1] Stéphan Suffit,[2] Lianhe Li,[3] Alexander Giles Davies,[3] Edmund Linfield,[3] Carlo Sirtori[1] and Yanko Todorov[1*]

[1]Laboratoire de Physique de l'Ecole Normale Supérieure, ENS, Paris Sciences et Lettres, CNRS, Sorbonne Université, Université de Paris, 24 Rue Lhomond, 75005 Paris, France

[2]Laboratoire Matériaux et Phénomènes Quantiques, Université de Paris, CNRS, 75013 Paris, France

[3]School of Electronics and Electrical Engineering, University of Leeds, LS2 9JT Leeds, United Kingdom

**We demonstrate a high temperature performance quantum detector of Terahertz (THz) radiation based on three-dimensional metamaterial. The metamaterial unit cell consists of an inductor-capacitor (LC) resonator laterally coupled with antenna elements. The absorbing region, consisting of semiconductor quantum wells is contained in the strongly ultra-subwavelength capacitors of the LC structure. The high radiation loss of the antenna allows strongly increased collection efficiency for the incident THz radiation, while the small effective volume of the LC resonator allows intense light-matter coupling with reduced electrical area. As a result our detectors operates at much higher temperatures than conventional quantum well detectors demonstrated so far.**

The terahertz (THz) domain of the electromagnetic spectrum has well defined applications ranging from imaging, security, spectroscopy [1], and has been recently envisioned for future 6G wireless communications by Samsung [2]. Quantum detectors based on electronic transitions in semiconductor quantum wells, QWIPs [3,4] have many assets for advanced applications, as they combine high speed and high quantum efficiency. Furthermore, they can be implemented in coherent schemes of detection [5, 6] and they can also be integrated in optical circuits alongside laser sources [7]. However, high temperature operation of THz QWIPs remains notoriously a difficult task. Indeed, typical THz photon energies are small (~10 meV, $\lambda$=100 µm) and thermally excited dark current quickly overrides the photocurrent signal when temperature is increased above 10 K [3, 4]. Here we propose a metamaterial-based concept which allows a significant increase of the detector operating temperature, up to 60 K. This is achieved because of the ability of our metamaterial to keep intense light-matter interaction in a highly subwavelength volume [8, 9]. As a result, the electrical area and the associated dark current noise are strongly suppressed [10, 11]. This metamaterial detector has the potential to reach liquid nitrogen operation, like commercial detectors in the $\lambda$=8-10 µm band. Furthermore, such devices that combine electrical read-out with enhanced light-matter interaction provide a technological background for quantum-optical experiments with THz photons [12].

In our device we combine quantum detector based on electronic transitions in quantum wells (THz QWIP, [3]) with a three-dimensional metamaterial architecture, as described in Figure 1(a). The unit cell of the metamaterial is a sub-wavelength inductor-capacitor (LC) resonant circuit [8, 9], scaled to operate at the frequency of the QWIP electronic transition, 5.2 THz. The LC resonator is further combined with half-wavelength ($\lambda$/2) patch antennae in order to boost the effective collection area of the device [9]. In addition, our metamaterial has been provided with electrical leads which allow an electrical bias on the QWIP and thus the extraction of a photocurrent signal. The QWIP absorbing region is contained only in the capacitive parts of the metamaterial, and contains only eight 15.5 nm GaAs quantum wells separated by 70 nm wide $Al_{0.03}Ga_{0.97}As$ barriers (Figure 1(b)). Each quantum well is nominally doped at $6x10^{10} cm^{-2}$. Highly doped, 40 nm wide n++ layers on both side of the absorbing region ensure ohmic contacts with the electrical leads. The entire QWIP region is thus $L_{tot}$ = 834 nm thick, in order to keep strong confinement between the quantum wells and the electric field of the LC resonators [8]. Figure 1(c) shows an optical microscope picture of the metamaterial array, indicating the array unit cell $\Sigma$ (dashed square). Typically, $\Sigma$=80 µm², while the overall optical surface of all devices studied here is 120x140 µm². The electrical area of each unit cell corresponds to the overall surface of the capacitor parts, $\sigma$ = 4.5 µm². The high value of the ratio $\Sigma/\sigma \sim 18$ ensures the high operating temperature of the detector [10]. Further details on the device fabrication are provided in the SI.

As indicated in Figure 1(d) each antenna connects two neighboring LC units, ensuring thus maximum possible packing of the array [9]. As discussed in Ref. [9], the LC resonator and the antenna are two strongly-coupled optical oscillators, with a typical coupling strength of $G$=0.8 THz. This is further illustrated in Figure 2, which provides an estimate of the optical response of the metamaterial based on coupled mode theory (see [9] for details). While the LC resonator alone is very poorly coupled to free space, the coupling with the antenna with high radiation loss allows a significant increase of the reflectivity contrast of the coupled system (Fig. 2(a)). At the same time, the small effective volume of the LC resonator allows intense light-matter coupling with reduced electrical area. The resonant frequency of the THz coupled-modes depends on two parameters: the perimeter of the LC inductive loop, $P$, and the antenna length $L_a$ (Fig. 1(d)). Here we report on three devices with $P$=4 µm and variable antenna length $L_a$=7 µm, 8 µm and 10 µm so that the mode at higher energy is close to the QWIP absorption resonance (Fig. 2(b)). In the SI we provide an analysis of the reflectivity curves.

Alongside metamaterial detectors we also studied a reference device where the absorbing region was processed into a square mesa of dimensions 200x200 µm² [10, 13, 14]. In this geometry light is coupled through a 45° wedge of the semiconductor substrate. Such device allows evaluating the performance of the detector region without any metamaterial effect, as well as obtaining the photoconductive gain [4].

We performed optical and electrical characterizations of the devices as a function of the temperature. In particular we compared the current – voltage characteristics of the mesa and of the metamaterial device in dark conditions, as well as under illumination from ambient black body radiation at 300K. In dark conditions, the sample is enclosed in a metallic cryo-shield without aperture that is maintained at the same temperature as the detector. For measurements under background illumination, the cryo-shield has an opening such as the field of view of the sample is ~60°. Figure 3 presents the results for the mesa (Figure 3(a)) and the metamaterial detector with $P$=4 µm, $L_a$=7 µm (Figure 3(b)). In this figure the background current density $J_B$ and the dark current density $J_{dark}$ are plotted as a function of the applied bias for different temperatures. For each device the current density has been calculated by considering the total electrical area. In both cases, we observe a dark current density that increases exponentially with the temperature. At fixed bias, the dark current is well fitted by the formula $J_{dark} = J_0 \exp(-E_{act}/k_B T) + const$ where the activation energy is $E_{act}$ = 13.2 meV (3.2 THz) for the mesa device and $E_{act}$ = 15.6 meV (3.8 THz) for the metamaterial device (see SI for fits). Furthermore, the dark current in the mesa is typically 10 times higher than that in the metamaterial device. This difference can be partly explained by the charge depletion of the devices processed in metamaterial geometry [8] (see further). In the case of the mesa, the background and the dark current-voltage characteristics become almost identical at 20 K. On the contrary, for the metamaterial structure, substantial difference between the background and the dark current remains up to 40 K. At *T*=4 K , the background current of the metamaterial is 2400 times higher than the dark current, whereas for the case of the mesa the corresponding factor is ~30. These observations already show qualitatively that the metamaterial geometry leads to a very strong suppression of the dark current with respect to the photocurrent. In the following, we will comment the device performance at low applied bias (-20mV), as at higher bias the structure becomes electrically unstable owe to field domain formation [14].

Figure 4 presents the spectral response and responsivity of the structures. In Fig. 4(a) we report the spectrally resolved measurements of the photocurrent with the help of a Fourier Transform Infrared Spectrometer in a step-scan mode. Remarkably, on sample (*P*=4 µm, $L_a$ =7 µm) a clear spectrum was measured up to 60K, a record high operation temperature of a THz quantum detector. For devices with different values of *P* and $L_a$ we routinely observed measurable spectral response up to 40K. Further characterizations as a function of the polarization of the incident wave were also performed, as shown in the SI.

The peak responsivity which corresponds to the maximum of the photocurrent spectra in Fig.4(a) is reported in Figure 4(b). It was obtained from the photocurrent measured by exposing the sample to a 500°C calibrated black-body source and using a lock-in detection technique with a chopper (120 Hz). Samples were mounted in the same conditions as for the background-limited experiments, and light from the blackbody was refocused on the sample with the help of two F1 off-axis parabolic mirrors (full details are provided in SI). For adequate estimation of the photon flux we also took into account the 300 K radiation from the shutters of the chopper. In Figure 4(b) we compare the resulting responsivity as a function of the detector temperature for the mesa and the (*P*=4 µm, $L_a$ =7 µm) detector at two different voltages, +20 mV and -20 mV. In the case of the mesa we were able to measure the responsivity only up to 20 K, as the signal at higher temperatures was overridden by dark current noise. In the metamaterial device a photocurrent signal was measured up to 60K. As seen from Figure 4(a), the responsivity depends strongly on the temperature for the metamaterial detector: it increases sharply as the temperature is increased from 4 K to 20 K, and then drops as the temperature is further increased. Such variations of the photoconductive response have been observed before [3], although in that case studies were performed on much shorter temperature range (up to 16 K). The responsivity rise at low temperature can be explained by activation of carrier injection from the n$^{++}$ contacts to the quantum wells [15], this phenomenon could also explain the strong asymmetry observed between positive and negative bias. Another phenomenon that contributes to the low temperature responsivity is the carrier depletion in the metamaterial device at low temperature due to the etched active region sidewalls; such phenomena was previously observed with similar structures [8]. Note that our metamaterial geometry allows exploring a much wider temperature range of the detector operation with respect to the existing literature, and clearly evidences the temperature variations of the responsivity of such THz QWIP. In the case of the mesa, we observe a maximum value of the responsivity 0.1 A/W at +20 mV, while the maximum value is 0.25 A/W for the metamaterial sample.

The external responsivity of all structures can be written in the form [4] $R = eg\eta/E_{21}$ where $e$ is the electron charge, $g$ the photoconductive gain (number of photo-excited electrons read-out by an external circuit per absorbed photon), $\eta$ is external quantum efficiency (number of photons absorbed by the intersubband transition with respect to the incident photon number) and $E_{21}$ is the energy of the electronic transition. In the case of the metamaterial resonators, $\eta$ depends on the properties of the LC and antenna elements as described in Ref. [9]. In the mesa geometry with a $\theta$=45° polished facet, $\eta_{mesa}$ can be evaluated from the well-known expression [13]: $\eta_{mesa} = (2\sin^2\theta/\cos\theta)Tf_{21}e^2N_{QW}N_s/(m^*nc\varepsilon_0\Gamma)$. Here $T = 0.67$ is the transmission of the facet, $f_{21}=0.7$ is the oscillator strength of the electronic transition, obtained from bandstructure simulations of the quantum wells (see also Figure 1(b)), $N_{QW} = 8$ is the number of quantum wells, $N_s = 6\times10^{10}$ cm$^{-2}$ is the areal electronic density per well, $m^* = 0.067m_0$ is the effective electron mass, $n=3.7$ is the refractive index at 5 THz, $c$ is the speed of light, $\varepsilon_0$ the vacuum permittivity and $\Gamma$=0.8 THz is the half width at half maximum of the electronic transition obtained from fits of the photocurrent spectra from Figure 4(b). In the expression of $\eta_{mesa}$ we also take into account the polarization response of the electronic system. In the present case we obtain $\eta_{mesa} = 0.026$, which provides a photoconductive gain $g$=0.1 at a bias of +20mV for a maximum value of the responsivity $R$=0.1 A/W. Similar values of the gain have been observed in previous studies [14]; however direct comparison between the samples is difficult owe to the different number of quantum wells, as well as the strong dependence of the gain on the applied bias [14]. Indeed, previous studies with QWIP operating at wavelengths $\lambda$=9 µm have shown that the gain in thin quantum well structures is very sensitive to the electric field inhomogeneity and injection from the contacts that lead to reduction of the responsivity [16].

In the case of metamaterial detectors, the photon absorption efficiency $\eta$ can be evaluated by systematic studies of the reflectivity spectra of the arrays and by applying coupled-mode theory equations [9]. In practice, it was difficult to evaluate experimentally the reflectivity of the detectors because of their small area. Furthermore, the exact evaluation of $\eta$ is complicated by the charge depletion phenomenon discussed above. However, we could evaluate $\eta$ from the measurements of the responsivity and assuming that the mesa and metamaterial have the same gain at the maximum responsivity point. The corresponding values of $\eta$ estimated for the metamaterial devices are $\eta = 0.06$ ($P = 4$ µm, $L_a = 7$ µm), $\eta = 0.034$ ($P = 4$ µm, $L_a = 8$ µm) and $\eta = 0.04$ ($P = 4$ µm, $L_a = 10$ µm). While these values are superior to the mesa reference ($\eta_{mesa} = 0.026$), they are of the same order of magnitude. Indeed, the superior temperature performance of the metamaterial arises from the strong reduction of the dark current noise, as discussed further.

The background limited detectivity of the structure has the following expression [4, 10, 17]:

(1) $$D^*_{BLIP} = \sqrt{\frac{\Sigma}{\sigma}}\frac{R}{\sqrt{4egJ_B}}$$

This expression reveals the effect of metamaterial detectors through the factor $(\Sigma/\sigma)^{1/2}$ [10]: in the present designs we have $\Sigma/\sigma = 14-21$. This brings an enhancement of a factor of 4 of the high-temperature detectivity with respect to the mesa geometry. Note that here we are interested in particular to the "high temperature" regime where the detector is dominated by the dark current noise $J_B\sim J_{dark}$. The dependence of $D_{BLIP}^*$ on $T$ resulting from equation (1) is shown in Figure 5. For comparison we also plot the value $D_{max}=2.3 \times 10^{11}$ cmHz$^{0.5}$/W for an ideal photoconductive detector with unity quantum efficiency [17].

We observe that, in comparison with the mesa device, the enhanced detectivity compensates the exponentially increasing dark current, and thus expands the temperature operation of the device by about 40 K. The $P$ = 4 μm $L_a$ = 8 μm device where the carrier depletion was less pronounced has a maximum detectivity $D^*$ = 0.6x10$^{11}$ cmHz$^{0.5}$/W, at 15 K, an order of magnitude higher than the value of the reference mesa at 4K. This value corresponds to a Noise Equivalent Power, NEP = 0.2 pW. For the best high temperature performing metamaterial ($P$= 4 μm, $L_a$ =7 μm), we estimate a NEP = 150 pW at 60 K. The high temperature performance of our detector is thus comparable with recently demonstrated graphene based detectors [18], however with the potential to operate at much higher speed [19].

In conclusion, we have shown that our metamaterial design has a strong impact on the performance of THz detectors. In this first demonstration we showed that such devices can operate up to 60 K. However, both the metamaterial structure and the absorbing region dispose of many degrees of freedom to further improve the detector temperature operation. For instance we can envision quantum cascade detectors that operate at 0 V and feature even lower dark current [20]. Clearly, an optimization of the electronic density is also required to fully exploit the potential of the metamaterial geometry. We believe that an operation at 80 K is within reach: this would be a significant technological leap, as it would allow the use of compact and portable cryocoolers that are currently used at industrial scale. Also, the circuit-like nature of our metamaterial resonators allows its integration with other active devices in THz optical circuits. Finally, we have demonstrated that our metamaterial design, that was previously used to demonstrate ultra-strong light-matter coupling is also compatible with electrical leads. This makes it an interesting technological platform to explore the interplay between electrical transport and quantum-optical phenomena [12, 21].

**Supplementary material**

See supplementary material for full details on the sample fabrication, biasing scheme for the mesa and metamaterial detectors, method to estimate the reflectivity curves shown in Fig. 2, dark current fits, methods for measuring the detector responsivity, polarization dependence of the photo-response, as well as additional current voltage characteristics for the metamaterial devices commented in Fig. 5.

**Acknowledgments**

This work was supported by the French National Research Agency under the contract ANR-16-CE24-0020, the regional DIM-SIRTEQ project, "CIEL", and the Engineering and Physical Sciences Research Council (UK) grant EP/P021859/1. EHL acknowledges the support of the Royal Society and Wolfson Foundation.

**Data availability statement:** The data that supports the findings of this study are available within the article [and its supplementary material].

**Corresponding author:** yanko.todorov@phys.ens.fr

ACCEPTED MANUSCRIPT

Applied Physics Letters

This is the author's peer reviewed, accepted manuscript. However, the online version of record will be different from this version once it has been copyedited and typeset.

PLEASE CITE THIS ARTICLE AS DOI: 10.1063/5.0033367

**Figures**

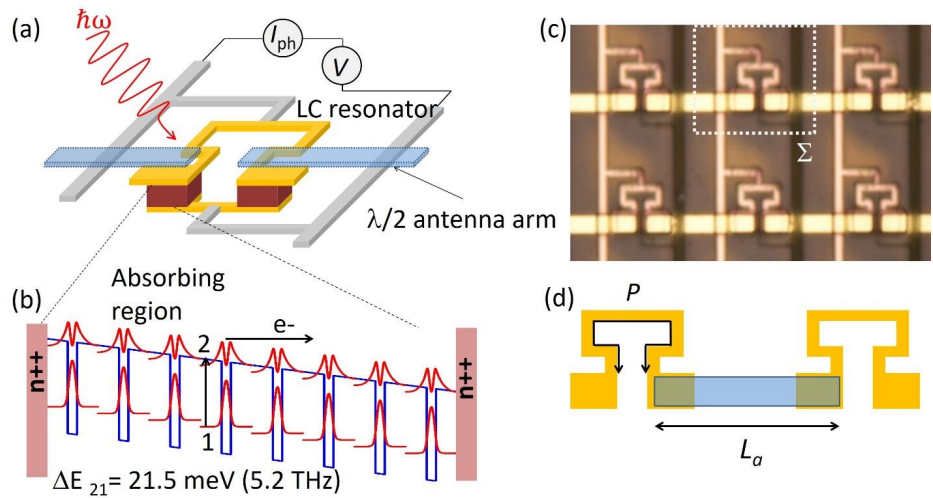

**Figure 1.** (a) Schematics of the metamaterial unit cell, which comprises a three dimensional LC resonator laterally coupled with antennas. The inductive parts of the LC resonator have leads which allow application of a bias as well as read-out of photocurrent signal. The photoconductive region is contained in the capacitive parts of the LC circuit. (b) Details of the absorbing region, which contains eight quantum wells with an electronic transition at 5.2 THz. (c) Optical microscope picture of the metamaterial. Each antenna connects two LC resonators. The dashed rectangle corresponds to the metamaterial unit cell $\Sigma$. (d) Schematics of two LC resonators connected by an antenna. The relevant parameters that set the metamaterial resonance are the inductive loop length $P$ and the length of the antenna $L_a$.

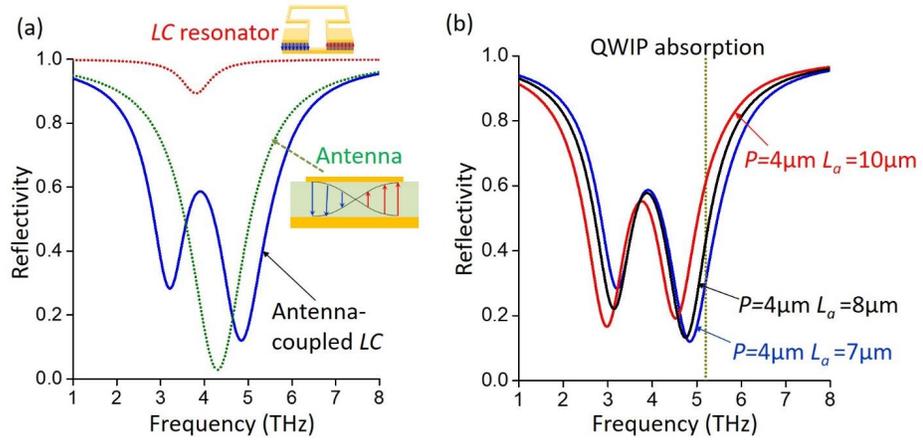

**Figure 2.** (a) Illustration of the strong coupling between the LC resonators and the antenna resonance. The dotted curve show simulations of the reflectivity curves of the uncoupled resonators, which are schematized in the inset images. The LC resonator has a poor coupling with free space (low reflectivity contrast), while the antenna has strong radiation loss (almost unity contrast). The two coupled modes (continuous line) inherit the high radiation loss of the antenna, while keeping the strong electromagnetic confinement of the LC resonator. (b) Simulation of reflectivity curves of the metamaterials where the high energy coupled mode is close to the QWIP absorption resonance.

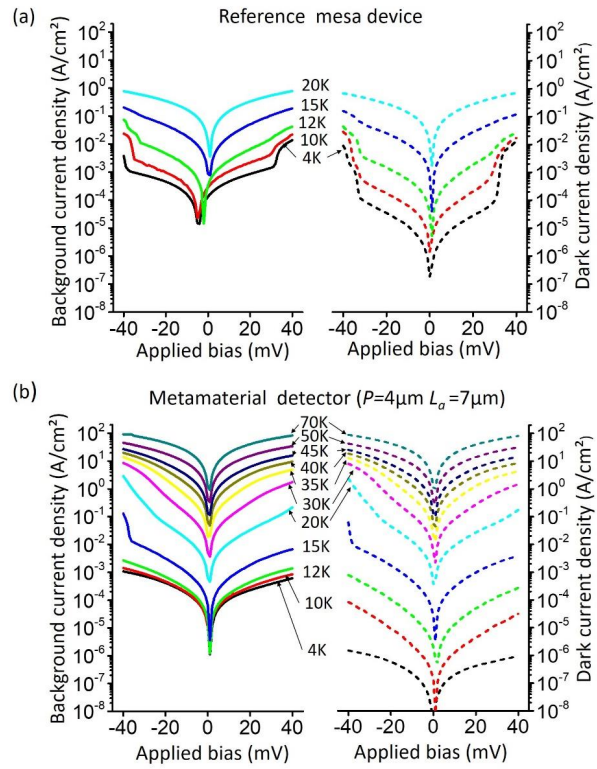

**Figure 3.** (a) Current-voltage characteristics of the mesa reference device, as a function of the device heat-sink temperature. The dotted curves are dark current density, whereas the continuous lines are background current where the detector is exposed to the ambient thermal radiation. (b) Current-voltage characteristics of a metamaterial device, with the same convention as in (a).

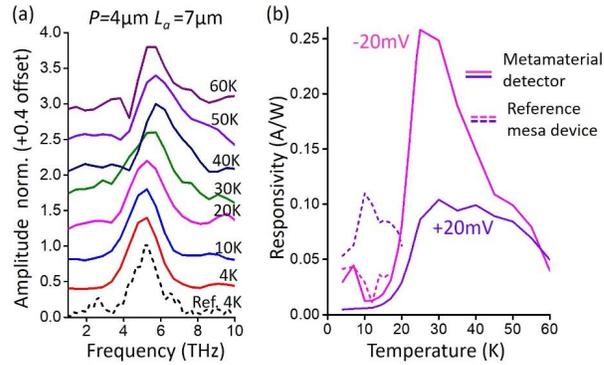

**Figure 4.** (a) Photocurrent spectra at different temperatures for the metamaterial detector. The dashed curve reports the mesa spectrum at 4K. (b) Responsivity as a function of the detector temperature at bias +20mV (purple) and -20mV (magenta). The dashed curves are the results for the reference mesa device, and the full lines represent the metamaterial detector with $P$= 4μm and $L_a$= 7μm.

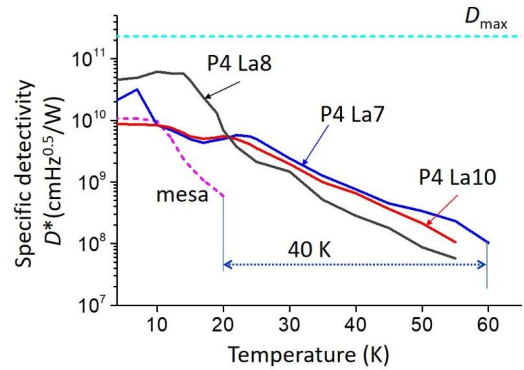

**Figure 5.** Estimates of the detectivity as a function of the temperature for an applied bias of -20mV on the metamaterial arrays and +20mV for the mesa device. $D_{max}$=2.3x10$^{11}$cmHz$^{0.5}$/W corresponds to the maximum possible background limited detectivity for a perfect photo-conductive detector.

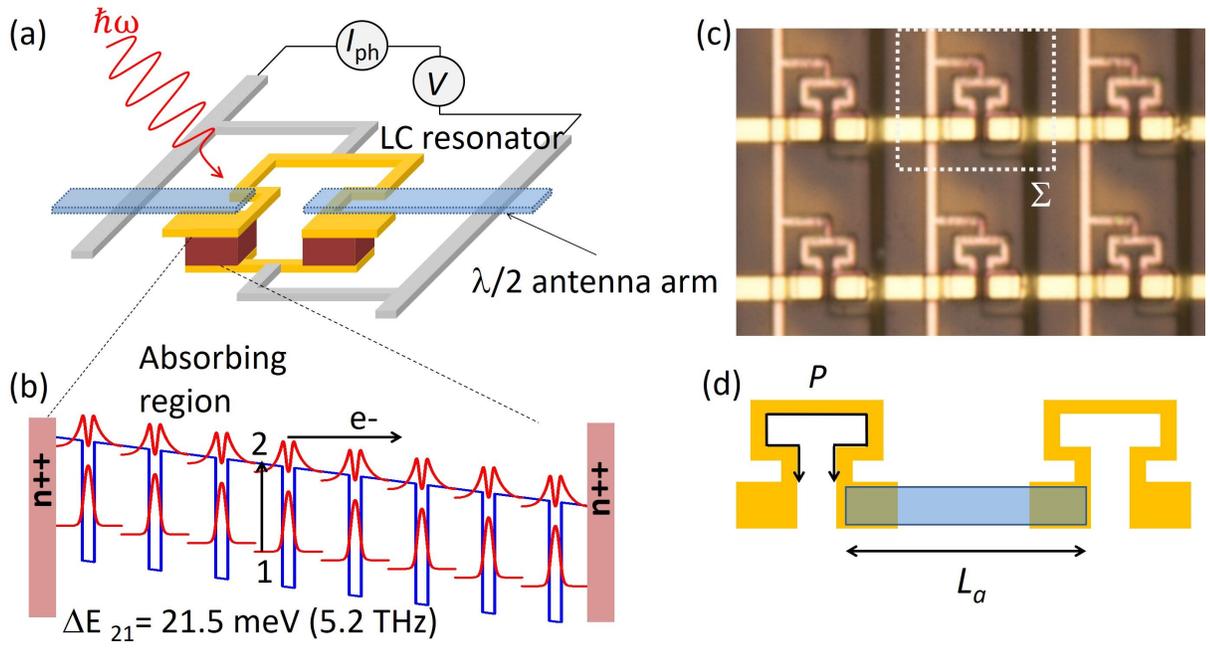

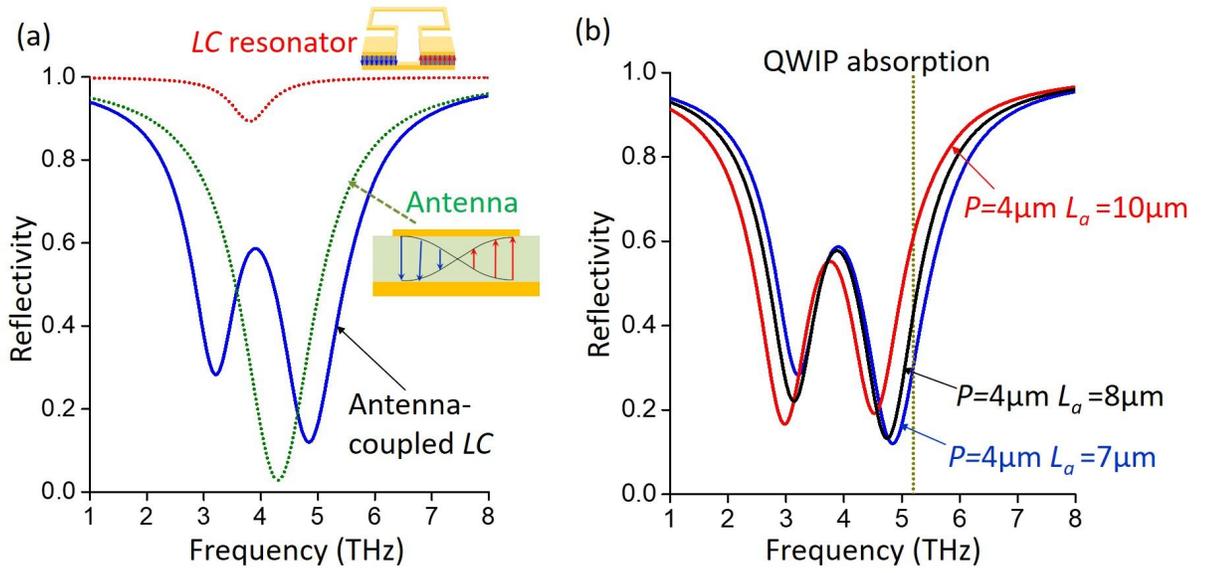

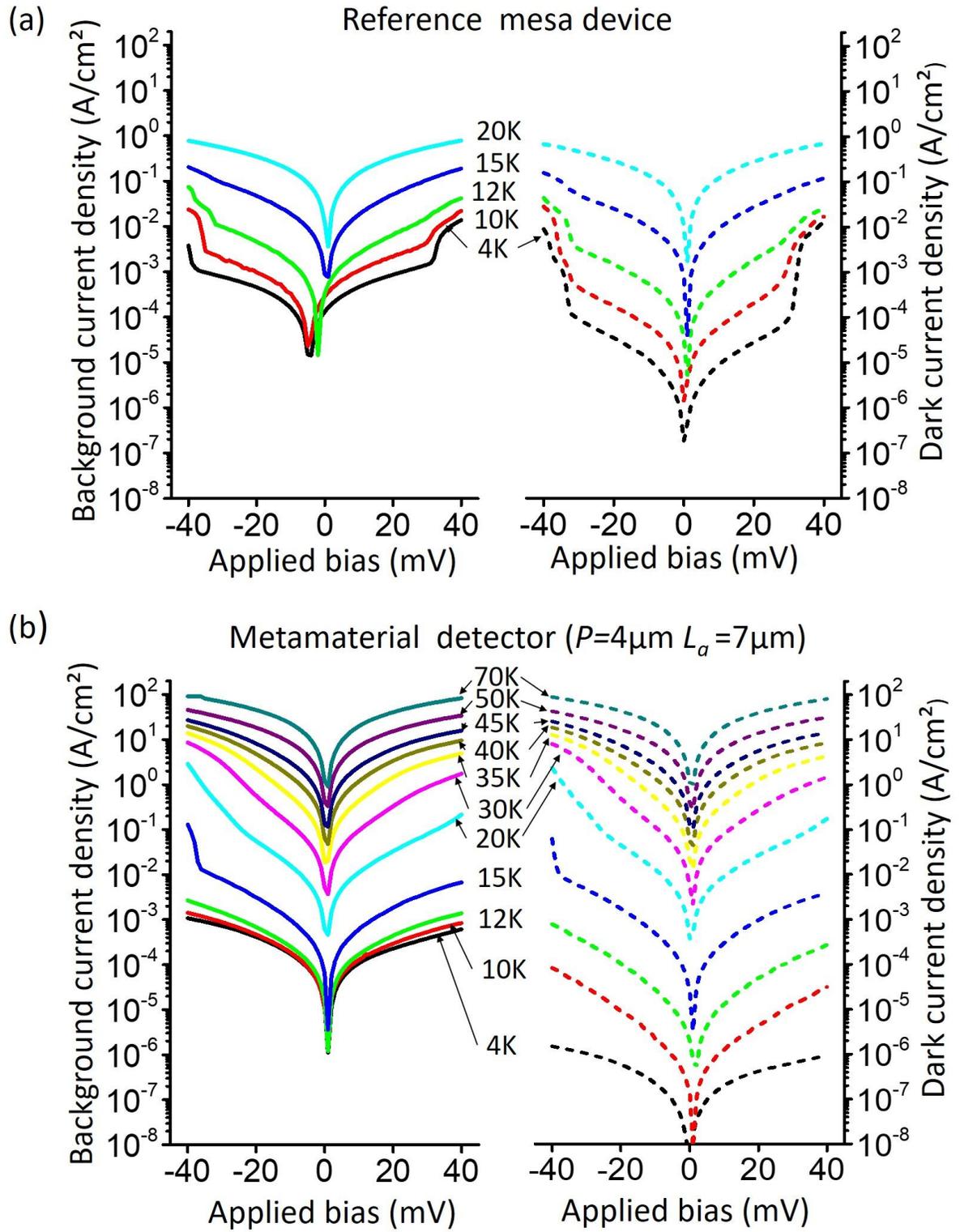

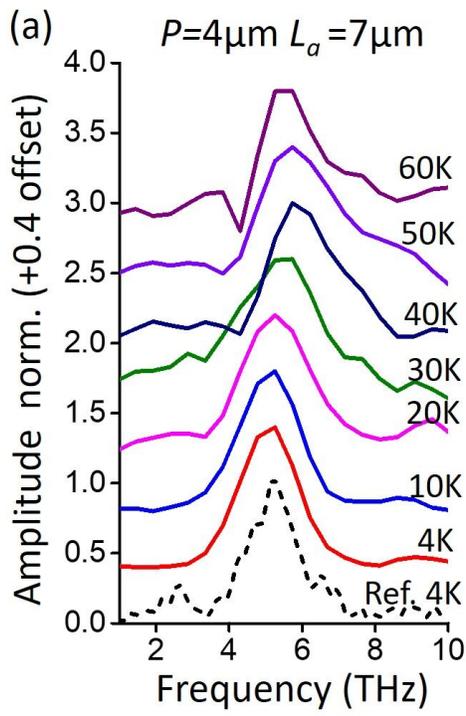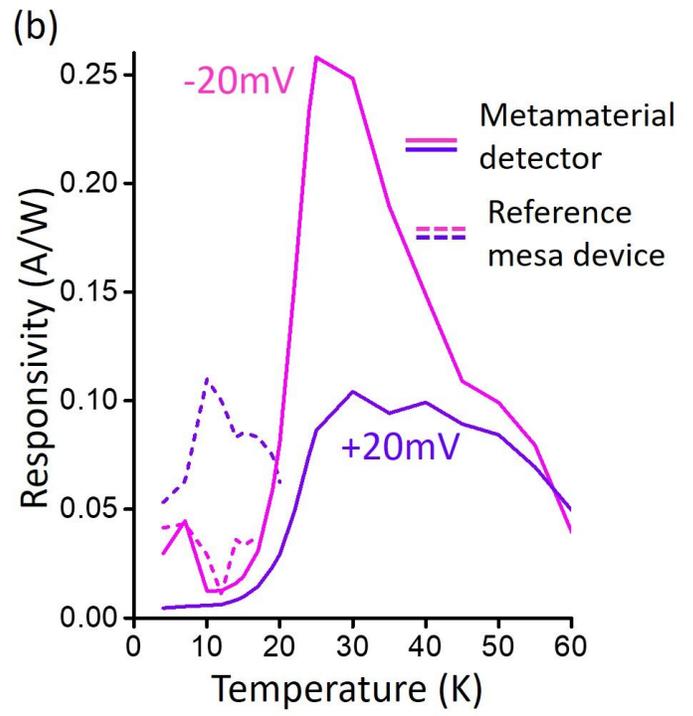

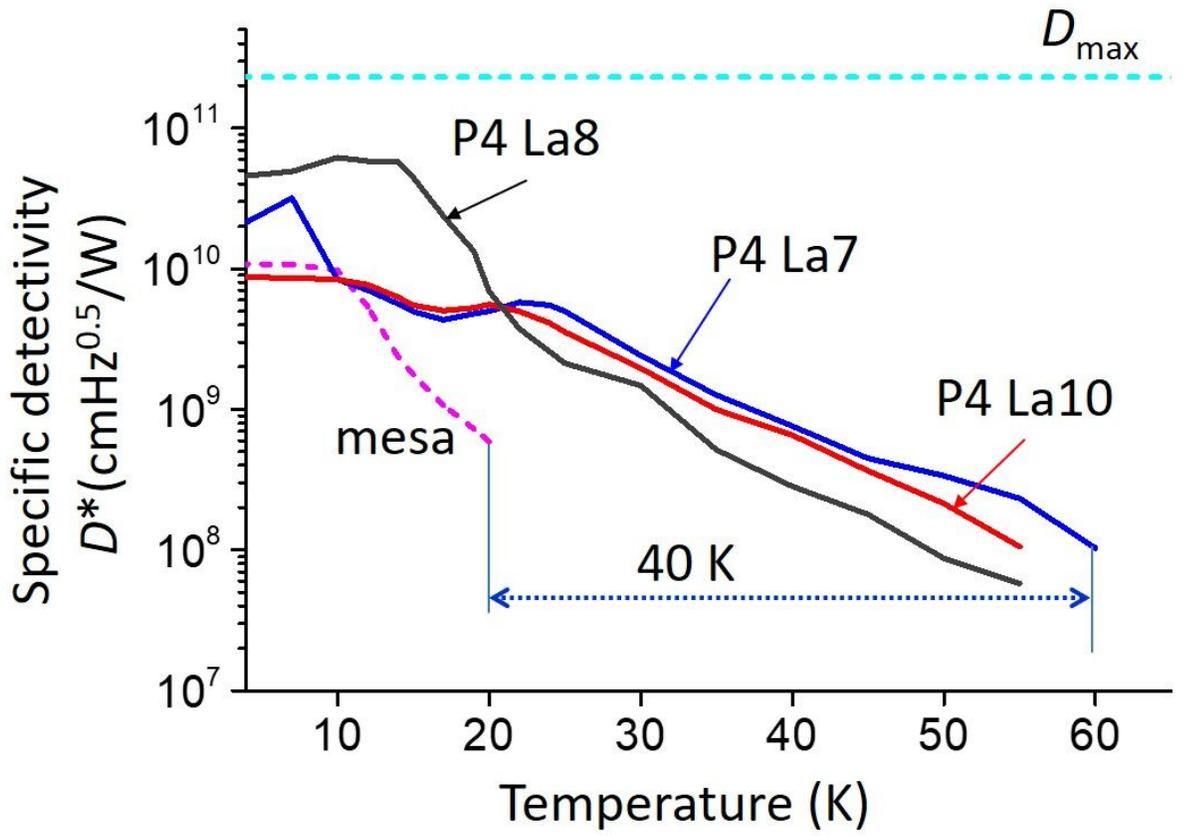

# Supplementary Information

# High temperature metamaterial TeraHertz quantum detector

Mathieu Jeannin,[1] Thomas Bonazzi,[1] Djamal Gacemi,[1] Angela Vasanelli,[1] Stéphan Suffit,[2] Lianhe Li,[3] Alexander Giles Davies,[3] E dmund Linfield,[3] Carlo Sirtori[1] and Yanko Todorov[1*]

[1]Laboratoire de Physique de l'Ecole Normale Supérieure, ENS, Paris Sciences et Lettres, CNRS, Université de Paris, 24 Rue Lhomond, 75005 Paris, France

[2]Laboratoire Matériaux et Phénomènes Quantiques, Université de Paris, CNRS, 75013 Paris, France

[3]School of Electronics and Electrical Engineering, University of Leeds, LS2 9JT Leeds, United Kingdom

1. Sample fabrication

The fabrication procedure closely resembles the one described in Refs. [1, 2], but has been adapted to improve the optical quality of the sample while coping with the high temperature annealings necessary to obtain ohmic contacts between the active region and the metallic parts.

We first fabricate the *LC* resonators. We start by fabricating the 1.5x1.5 µm² ohmic contact pads below the bottom (square) capacitor plates of the LC circuits using electron beam lithography (EBL) and electron beam evaporation (EBE) of a stack of Pd/Ge/Ti/Au (15/45/5/15 nm). Using a subsequent EBL step, we pattern the bottom part of the LC circuit consisting of two capacitor plates on top of the ohmic contacts, and the straight wire connecting the two pads. We also pattern the metallic leads (long vertical wires in Fig. 1 of the main text) and evaporate a Ti/Au (5/150 nm) stack. Large contacting pads overlapping the vertical metallic leads are added using optical lithography to allow contacting the bottom side of individual arrays using wirebonding. The sample is then etched using the metal as a mask using a Cl-based inductively coupled plasma reactive ion etching (ICP-RIE). A 3 µm thick SiN insulation layer is deposited using a low temperature (150°C) plasma enhanced chemical vapor deposition (LT-PECVD) and coated with Ti/Au (10/200 nm). The sample is then flipped and bonded to a host GaAs substrate using an epoxy. The substrate is selectively etched, revealing the unprocessed side of the active region. We repeat the same two-step EBL+EBE procedure to fabricate the top contacts of the LC circuits (1.5x1.5 µm² Pd/Ge/Ti/Au pads) and then the top part of the LC circuit (Ti/Au capacitor plates, inductive loop and metallic leads). A second contacting pad is processed using optical lithography to allow contacting the top side of individual arrays. The excess active region is then removed using ICP-RIE again, revealing the metallic wires and pads buried below the active region. After this etch step, only the active region embedded in the capacitors of the LC circuits is left in the sample. Finally, a 500nm thick SiN layer is added to the sample using LT-PECVD, and the λ/2 antennas are patterned using EBL and deposited using EBE (Ti/Au 5/150nm). The SiN layer above the contact pads is then selectively removed using optical lithography and fluoride based reactive ion etching. Ultimately, the sample is annealed at 200°C on a hotplate for 30 min.

We summarize below the electrical area of each device and compare it to the physical footprint on the sample.

|  | La = 7 µm | La = 8 µm | La = 9 µm | La = 10 µm | La = 11 µm |
|---|---|---|---|---|---|
| Footprint (µm²) | 2.32 10⁴ | 2.24 10⁴ | 2.16 10⁴ | 2.08 10⁴ | 2 10⁴ |
| P = 9 µm | **2916** | **2430** | **2268** | **1944** | **1782** |
| P = 8 µm | **3098.25** | **2581.875** | **2409.75** | **2065.5** | **1893.375** |
| P = 4 µm | **2754** | **2409.75** | **2237.625** | **2065.5** | **1721.25** |
| P = 3 µm | **3280.5** | **2733.75** | **2551.5** | **2187** | **2004.75** |
| P = 2.4 µm | **3280.5** | **2916** | **2551.5** | **2187** | **2004.75** |

Table S1: Electrical area (bold numbers, in µm²) of each array, and the corresponding footprint on the sample (second line).

For the reference mesa device we use standard fabrication protocol, where a top PdGe contact is defined as above. The mesa is then defined by wet etching down to the bottom n++ contact, and then we evaporate AuGeNi contacts laterally to the mesa. The lateral contacts are diffused in order to reach the bottom n++ layer.

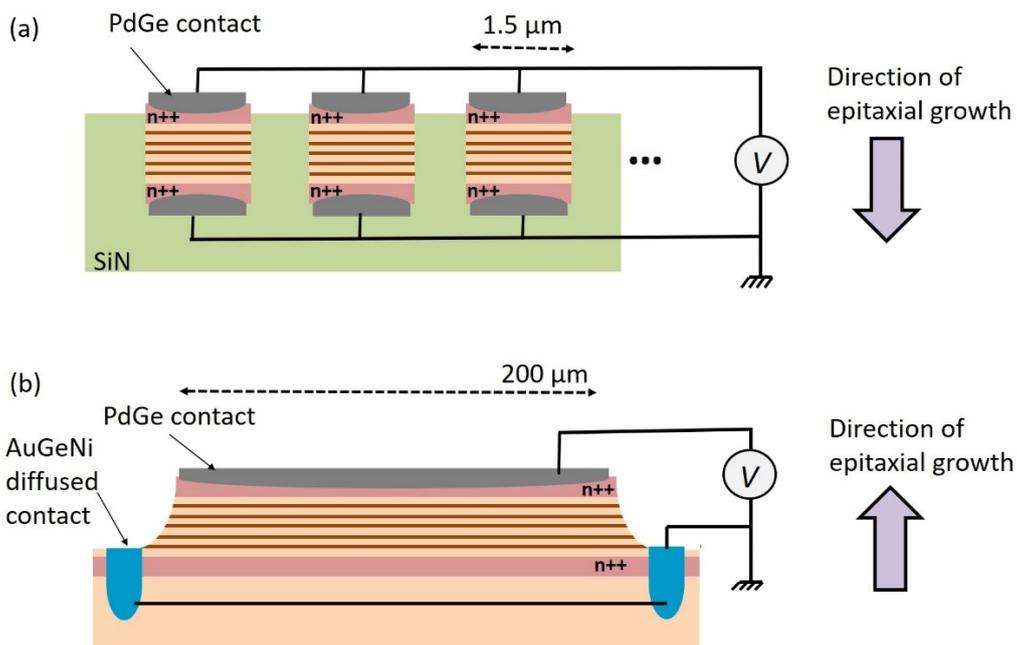

**Figure S1:** Illustration of the biasing scheme for the metamaterial device (a) and the mesa device (b) with respect to direction of the epitaxial growth. These illustrations are not on scale.

In Figure S1 we illustrate the electrical bias scheme for both the metamaterial and mesa device as a result of this process. Note that the as a result of the process, the absorbing region is inverted in the case of metamaterial devices with respect to the mesa device; we therefore compare the two devices for opposite sign of the applied bias. This comparison allows to take into account the asymmetries introduced by the epitaxial growth, such as dopant segregation [3].

## 2. Reflectivity curves

In Ref. [2] we have discussed in details the coupling between the LC resonators and the antenna elements. This coupling is described in terms of the temporal coupled mode theory [4], where the LC resonator and the antenna are treated as two coupled oscillators described by their frequencies, radiative and non-radiative loss rates: respectively ($\omega_{LC}$, $\Gamma_{LC}$, $\gamma_{LC}$) and ($\omega_A$, $\Gamma_A$, $\gamma_A$). The coupled-mode theory equations read:

(S1) $$\frac{da}{dt} - (i\omega_{LC} - \gamma_{LC} - \Gamma_{LC})a - iGA = \sqrt{2\Gamma_{LC}}\, S_{in}$$

(S2) $$\frac{dA}{dt} - (i\omega_A - \Gamma_A - \gamma_A)A - iGa = \sqrt{2\Gamma_A}\, S_{in}$$

(S3) $$S_{out} = -S_{in} + \sqrt{2\Gamma_A}\, A + \sqrt{2\Gamma_{LC}}\, a$$

Here $a$ and $A$ are the amplitudes of the LC resonator and the antenna mode; $S_{in}$ and $S_{out}$ are respectively the amplitude of the incident and reflected wave. The LC-antenna coupling is described by the coupling constant $G$. By solving the system (S1-S3) for a harmonic regime where all amplitudes evolve as ~$e^{i\omega t}$, we obtain the reflectivity $R(\omega) = |S_{out}|^2/|S_{in}|^2$ as illustrated in Figure 2 in the main text.

The system (S1-S3) includes several parameters that have to be determined experimentally. For this we use our prior results from Ref. [4] which are supplemented with novel measurements. First, we have prepared large area arrays with an antenna length $L_a$=13µm and a variable perimeter $P$. The resulting reflectivity spectra at room temperature and their modelling through the system (S1-S3) are shown in Figure S2.

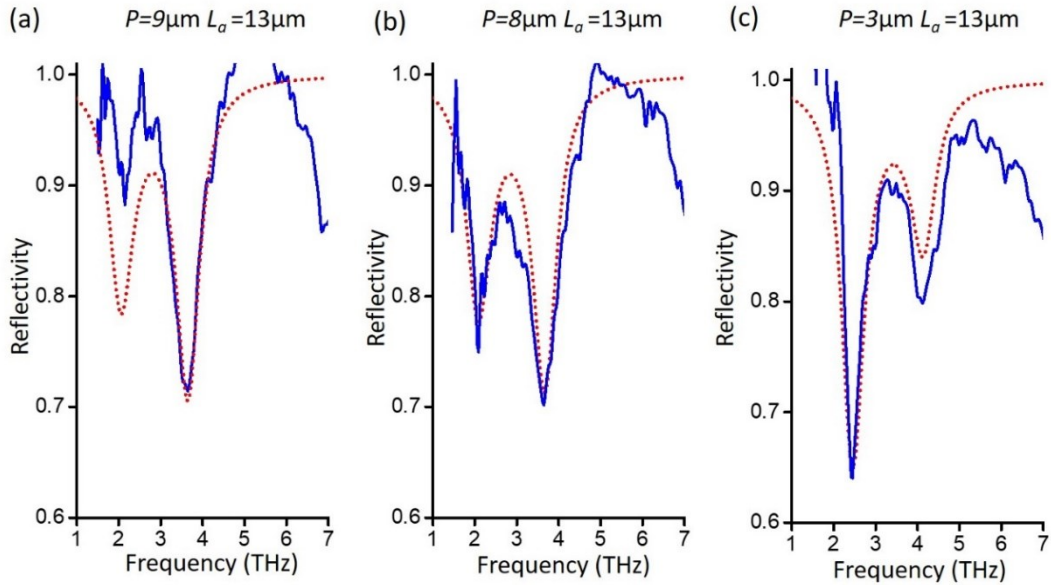

**Figure S2:** Reflectivity spectra at room temperature of large area (~1 mm²) arrays comprising $L_a$ = 13 µm long antennas and LC resonators with variable perimeter $P$: (*a*) $P$ = 9 µm, (*b*) $P$ = 8 µm and (*c*) $P$ = 3 µm. The full blue curves are experimental results and the dotted curves are modelling from the equations (S1-S3).

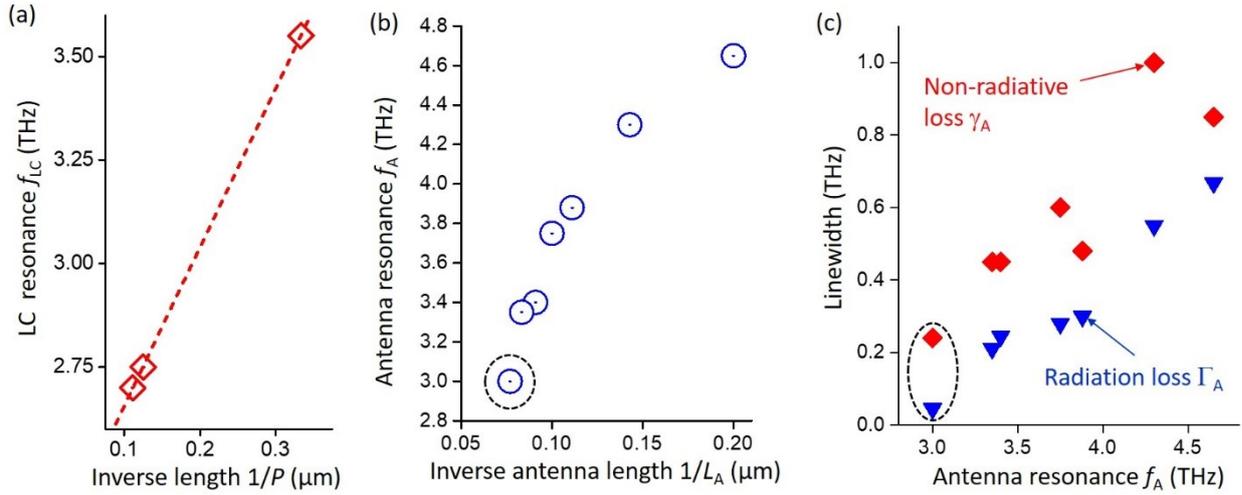

**Figure S3:** (a) LC resonant frequencies ($f_{LC}=\omega_{LC}/2\pi$, in THz) as a function of the inverse perimeter (1/$P$, in µm), as extracted from the data modelling of the data in Figure S2. (b) Resonant frequencies of the antennas ($f_A=\omega_A/2\pi$, in THz) obtained from the data in Ref. [2] and the data from Figure S1 (dot in dashed circle). (c) Non radiative ($\gamma_A$, diamonds) and radiative loss rates ($\Gamma_A$, triangles) obtained from Ref. [2] and the data from Figure S2 (symbols in dashed ellipse).

From these measurements we extract the various parameters such as the resonant frequencies of the uncoupled LC resonator (Figure S3(a)), as well as the magnitude of the coupling parameter $G$ = -0.8 THz. As in Ref. [2] we found that the radiation loss $\Gamma_{LC}$ of the LC structures are negligible, and they have non-radiative linewidth of $\gamma_{LC}$=0.4 THz. By studying uncoupled LC resonators alone, we found that the annealing of the ohmic contacts introduces a frequency shift $\Delta f_{LC}$=+0.56 THz that is taken into account in the final modelling shown in Figure 2. For the $P$ = 4 µm we find thus that the resonant frequency of the LC resonator alone is $f_{LC}$ = 3.8 THz.

The fit parameters from the data of Figure S1 are combined with the results from Ref. [2] to obtain all characteristics of the wire antennas (Figure S3(b,c)). In particular, we found the following values: $f_A$ = 4.3 THz for $L_a$=7 µm, $f_A$ = 4.07 THz for $L_a$=8 µm, and $f_A$ = 3.7 THz for $L_a$=10 µm. For all antennas we use the parameters $\gamma_a$=0.45 THz and $\Gamma_A$= 0.32 THz.

## 3. Dark current fits

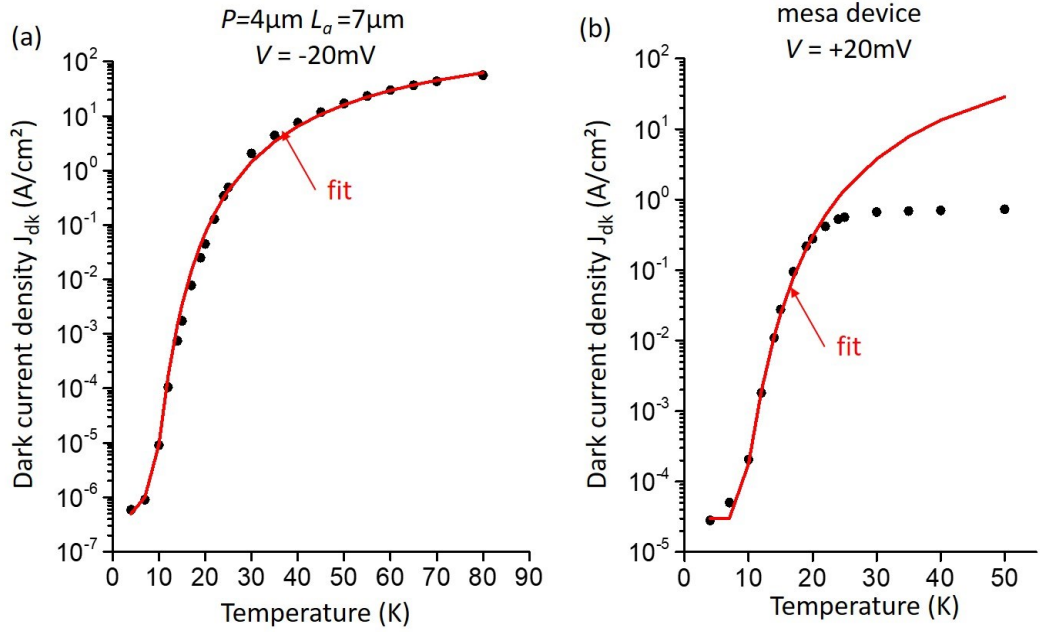

**Figure S4:** (a) Dark current density as a function of the temperature for the metamaterial device *P*= 4 μm and $L_a$ = 7 μm biased at -20 mV. The dots are experimental data and the full line is exponential fit. (b) The corresponding data for the mesa device.

The temperature activated dark current is fitted with the function $J_{dark} = J_0 \exp(-E_{act}/k_B T) + const$, as shown in Figure S4. The parameters for the metamaterial device (a) are $J_0$ = 2.8 A/cm², $E_{act}$ = 15.6 meV and *const* = 5.9x10⁻⁷ A/cm². For the mesa the corresponding parameters are: $J_0$ = 600 A/cm², $E_{act}$ = 13.1 meV and *const* = 3x10⁻⁵ A/cm². For the mesa, the deviation at high temperature can be explained by an extra resistance R ~ 10kΩ in series with the device $R_{device}$, such as the total resistance is R+$R_{device}$. At low temperature, the device is very resistive, $R_{device}$ ~ 100 MΩ, meaning that the effect of R is negligeable. At high temperature, we expect the device to have an intrinsic resistance on the order of $R_{device}$ ~ 1Ω, and therefore we see essentially the effect of *R*. The most probable explanation for the presence of *R* in for the case of the mesa is as follows. As illustrated in Fig. S1(a) for the mesa devices we use lateral bottom contact, with very thin, 40 nm, contact layer. Because of the very thin contact layers over a very large area mesa 200 μm x 200 μm we could expect an extra lateral resistance *R* in series with the resistance QW region. Note that we do not observe current saturation for the metamaterial detectors, even if the high temperature resistance is $R_{device}$ ~ 10 Ω. Indeed, in the case of metamaterials the QW are contacted with metal contacts on both sides, and furthermore the lateral sizes of the pads are 1.5 μm, which removes completely any effect of any lateral resistance (Fig. S1 (a)).

## 4. Photocurrent and responsivity measurements

In order to determine the responsivity of our devices we have used calibrated black body source heated at 500°C and a lock-in amplifier. Details of the experimental set-up are provided in Figure S5 (a). The blackbody radiation is focused on the sample with a help of a F1 parabolic mirror, after passing on a 50% mechanical chopper rotating at a speed 120 Hz. The current from the sample is amplified with a low-noise transimpedance amplifier and then send to the lock-in amplifier. The photocurrent $I_{ph}$ is related to the lock-in reading $V_{lock-in}$ through the formula $I_{ph} = V_{lock-in} (\pi/2^{1/2})/g_{tr}$, where $g_{tr}$ is the transimpedance gain.

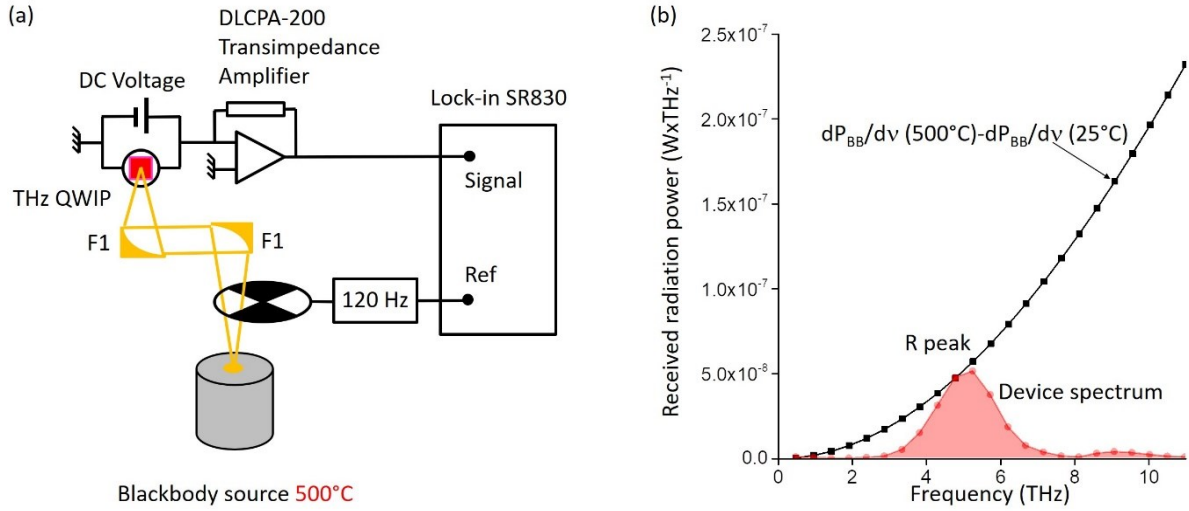

**Figure S5**: (a) Full experimental set-up for the measurements of the device responsivity. (b) Estimate of the received radiation power on the device. The black curve is a difference between 500°C and 25°C blackbody sources. The device spectrum has been normalized such as its maximum is set on the black curve. The resulting curve corresponds exactly to the integrand in Eq. (S5).

As we use a broadband source the measured photocurrent $I_{ph}$ corresponds to total signal collected from the entire spectral range of our THz detector. The spectral shape of the responsivity has been determined from spectrally resolved photocurrent measurements with the help of a FTIR as described in the main text and Figure 4(a). In order to obtain the correct spectral shape of the responsivity $R(\nu)$ the photocurrent spectra have been normalized on the power spectral density of the globar source of the FTIR assumed to be a blackbody of a temperature $T = 1200K$.

Knowing the spectral shape $R(\nu)$ responsivity we can determine the absolute value of the peak responsivity $R_{peak}$. Responsivity can be cast in the form:

(S4) $$R(\nu) = R_{peak} L(\nu)$$

Here $R_{peak}$ is the maximum responsivity at a frequency $\nu_{peak}$ = 5.2 THz (see Figure 4(a).) and $L(\nu)$ is the function that describes the responsivity spectrum but normalized such as $L(\nu=\nu_{peak})$ = 1. The total photocurrent measured from the lock-in experiment described above is thus provided by the formula:

(S5) $$I_{ph} = R_{peak} \int_0^\infty L(\nu) \left\{ \frac{dP_{BB}}{d\nu}(T=500°C) - \frac{dP_{BB}}{d\nu}(T=25°C) \right\} d\nu$$

Here $dP_{BB}/d\nu(T)$ is the spectral density of a black-body radiation incident on the detector that has been computed taking into account the device area and the solid angle of the F1 focusing mirror (see, for instance, equation (2.6) on page 7 from Ref. [5]). In this formula we take into account the fact that when the detector is facing the palette of the mechanical chopper it is actually exposed to the ambient black body radiation at 300K (~25°C). The function under the integral of Eq.(S5) has been plotted in Figure S5(b) together with the expression in brackets. The total power received by the device corresponds to the red area in the graph of Figure S5(b). For instance, for metamaterial devices we have typically an incident power 220nW. From this value we determine the peak responsivities shown in Figure 4(b) in the main text.

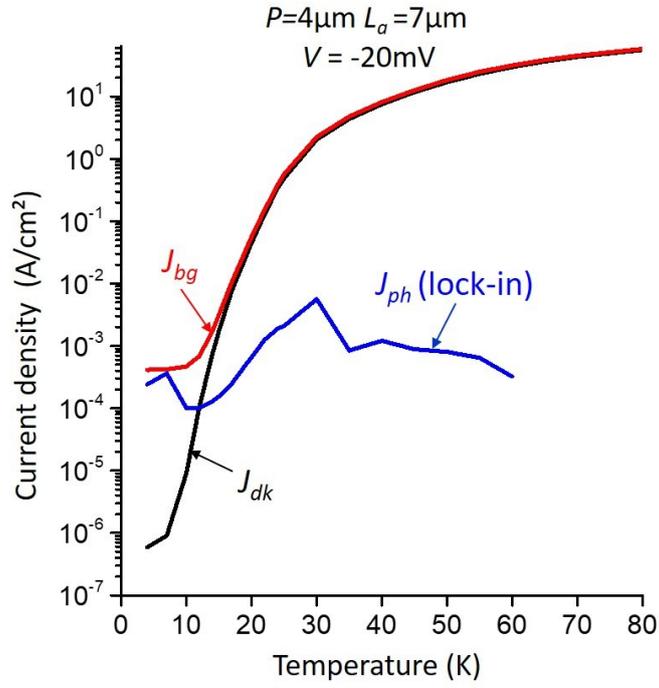

**Figure S6:** Comparison between current densities obtained from direct measurement of current voltage characteristics and photocurrent density measured in the lock-in experiment.

The DC measurements shown in Figure 2 of the main text were obtained with Yokogawa GS200 source/ measurement unit. In Figure S6 we show a comparison between DC current voltage characteristics and photocurrent obtained by a lock-in experiment described above. At low temperature, where the dark current is negligible we see that the background current is of the same order of magnitude as the photocurrent obtained from with the lock-in amplifier. These observations attest the reliability of our estimation for the responsivity.

5. **Polarization dependence of the photoresponse**

By construction the optical response of our metamaterial is dependent on the polarization of the incident THz wave. Only the electric field component of the incident wave, that is parallel to the wire antenna excites the metamaterial resonance [2]. As a consequence, the photoresponse of the detector is also dependent on the polarization of the incident polarization. This selection has been experimentally verified by placing a wiregrid THz polarizer in front of the detector, and measuring the spectral response as a function of the angle $\theta$ between the active polarization and the wire antennas of the metamaterial. These results are shown in Figure S7(a,b). As seen from Fig. S7(a) the photocurrent intensity is maximal when the incident electric field vector is aligned with the wire antennas. Furthermore, as shown in Fig. S7(b) the photocurrent signal follows the expected

cos²$\theta$ dependence. These results attest for the fact that the photoresponse of the device is mediated by the coupling between the metamaterial resonance and the electronic transitions, and not by any other spurious effects (i.e. bolometric effect). These measurements were performed with the ($P$=4μm, $L_a$ =7μm) device.

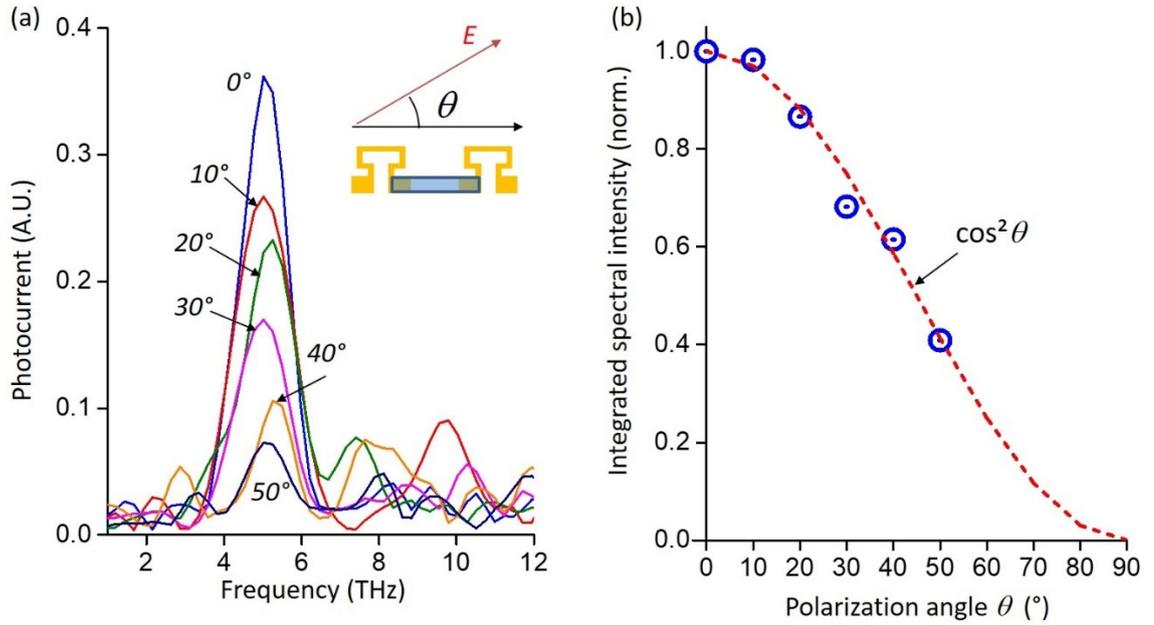

**Figure S7:** (a) Photocurrent spectra as a function of the polarization of the incident wave. The inset shows the electric field vector with respect to the wire antenna. In our experiments we vary the angle $\theta$ defined in the inset. (b) Values of the integrals of the photocurrent spectra as a function of $\theta$ (dots). The dashed line is a plot of the expected cos²$\theta$ dependence.

6. **Current-voltage characteristics of all structures reported in the main text**

Here we provide the current voltage characteristics of the structures ($P$ = 4 μm, $L_a$ = 8 μm) and ($P$ =4 μm, $L_a$= 10μm) that were used to obtain the results reported in Figure 5.

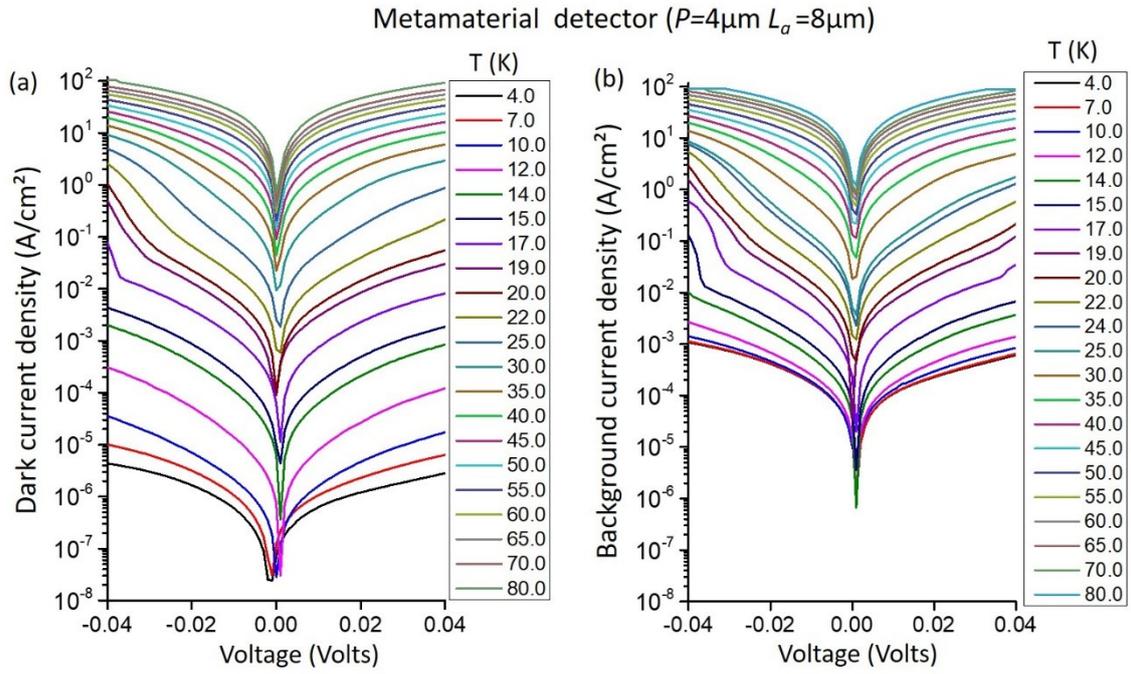

**Figure S8:** Current-voltage characteristics for under dark conditions (a) and background illumination (b) for the sample ($P$=4μm, $L_a$=8μm).

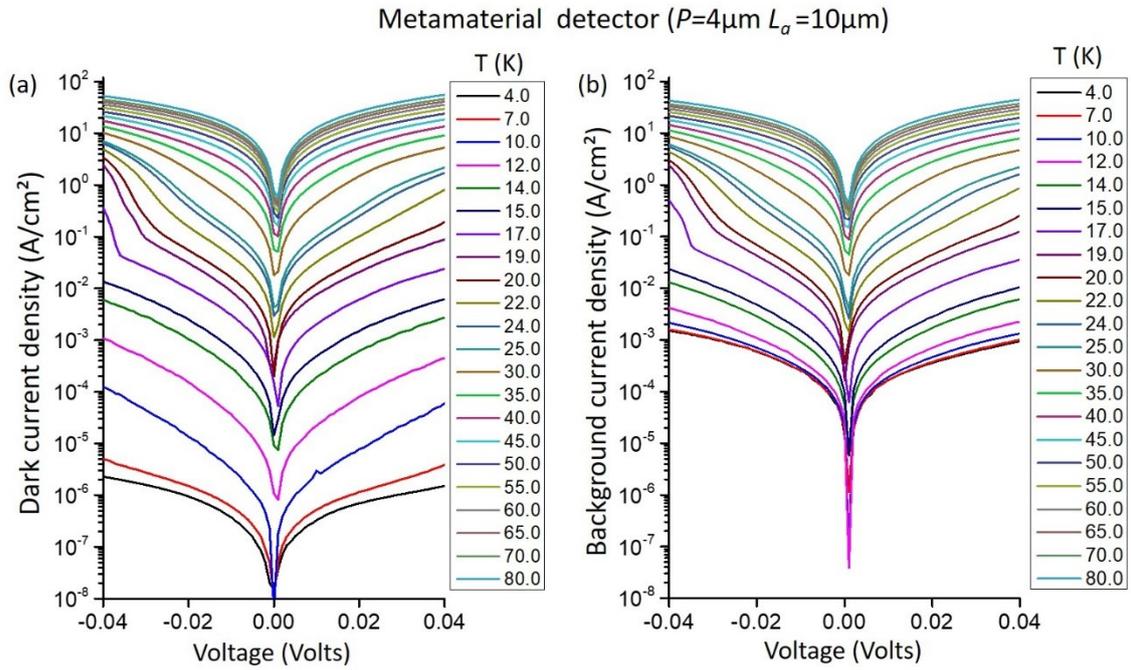

**Figure S9:** Current-voltage characteristics for under dark conditions (a) and background illumination (b) for the sample ($P$=4μm, $L_a$=10μm).